\documentclass[11pt]{article}
\pdfoutput=1
\usepackage{jheppub,color}
\usepackage{graphicx}
\usepackage{subfig}

\newcommand\beq{\begin{equation}}
\newcommand\eeq{\end{equation}}
\newcommand\be{\begin{equation}}
\newcommand\ee{\end{equation}}
\def\be{\begin{eqnarray}}
\def\ee{\end{eqnarray}}

\def\Dslash{\,\,{\raise.15ex\hbox{/}\mkern-12mu D}}
\def\Dbarslash{\,\,{\raise.15ex\hbox{/}\mkern-12mu {\bar D}}}
\def\delslash{\,\,{\raise.15ex\hbox{/}\mkern-9mu \partial}}
\def\delbarslash{\,\,{\raise.15ex\hbox{/}\mkern-9mu {\bar\partial}}}
\def\pslash{\,\,{\raise.15ex\hbox{/}\mkern-9mu p}}
\def\calDslash{\,\,{\raise.15ex\hbox{/}\mkern-12mu {\cal D}}}

\def\lae{\mathrel{\mathop{\smash{\lower .5 ex \hbox{$\stackrel<\sim$}}}}}
\def\lae{\mathrel{\mathop{\smash{\lower .5 ex \hbox{$\stackrel>\sim$}}}}}

\title{Embedding three-dimensional bosonization dualities into string theory}

\author[a]{Kristan Jensen}
\author[b]{ and Andreas Karch}
\affiliation[b]{Department of Physics, University of Washington, Seattle, Wa, 98195-1560, USA}
\affiliation[a]{Department of Physics and Astronomy, San Francisco State University, San Francisco, CA 94132 }

\preprint{\today}

\emailAdd{kristanj@sfsu.edu,akarch@uw.edu}

\abstract{We give simple string theory embeddings of several recently introduced dualities between 2+1-dimensional Chern-Simons matter theories using probe brane holography. Our construction is reliable in the limit of a large number of colors $N$ with fixed Chern-Simons level $k$.}

\begin{document}
\maketitle

\section{Introduction}

Dualities have long been one of our most important tools to understand strongly coupled systems. By rewriting the quantum dynamics in the right set of variables, the correct (and sometimes weakly coupled) low energy degrees of freedom can often be identified. Much progress along these lines has been made recently in relativistic field theories in 2+1 dimensions involving Chern-Simons (CS) gauge theories coupled to matter. Dualities between non-Abelian CS gauge theories based on $U(N)$ gauge groups in the large $N$ limit had been identified by equating the putative dual theories to one and the same ``holographic" description in terms of a classical higher spin theory in one dimension up \cite{Aharony:2011jz,Giombi:2011kc,Aharony:2012nh}. These dualities were termed 3d bosonization since they related a CS theory with fermionic to one with bosonic matter. At finite $N$ these dualities turn out to be much richer, as first described in \cite{Aharony:2015mjs} following~\cite{Radicevic:2015yla} and elaborated on in \cite{Hsin:2016blu}. Here one has to be careful to distinguish whether the gauge group is actually $U(N)$ or $SU(N)$. In the former case, the net flavor charge, ``baryon number" in the particle physics language, is gauged and so does not correspond to a global conserved charge. What is special in the case of 2+1 dimensions is however that the $U(1)\subset U(N)$ gauge factor itself gives a conserved global charge, ``monopole number.'' The corresponding current $j=\frac{*\text{tr}(F)}{2\pi}$ is conserved identically (that is, irrespective of the equations of motion) by the Bianchi identity for $F$. Bosonization dualities map monopole number to baryon number in the dual $SU(N)$ theory. The duality we focus on in this note is the one that equates
\beq \label{uni} SU(N)_{-k+N_f/2} \mbox{ with } N_f \mbox{ fermions } \quad \leftrightarrow \quad
U(k)_N \mbox{ with } N_f \mbox{ scalars \,.}
\eeq
This duality is believed to be valid for all $N_f \leq k$. (See~\cite{Komargodski:2017keh} for a proposal to extend the duality somewhat beyond this ``flavor bound.'') Here all matter is in the fundamental representation of the gauge group. The subscript on the gauge groups indicates the CS level. The scalar theory is accompanied with a quartic potential for the fundamental scalars, whose coefficient is tuned to a critical point, making them ``Wilson-Fisher" scalars. The Abelian $k=N=N_f=1$ case of this duality has been shown in \cite{Karch:2016sxi,Seiberg:2016gmd} (with closely related work in \cite{Murugan:2016zal}) to act as a seed duality from which one can derive a wide web of previously known dualities, such as ordinary particle/vortex duality, the fermionic particle/vortex duality of \cite{Son:2015xqa,ws,mv,mam}, as well as several new Abelian dualities \cite{Karch:2016aux,Hsin:2016blu}. More recently generalizations of the bosonization duality \eqref{uni} to symplectic and orthogonal groups have been constructed in \cite{Aharony:2016jvv,Metlitski:2016dht}:
\begin{eqnarray} \label{so} SO(N)_{-k+N_f/2} \mbox{ with } N_f \mbox{ real fermions } &\leftrightarrow& \quad
SO(k)_N \mbox{ with } N_f \mbox{ real scalars } \\
\label{sp}
USp(2N)_{-k+N_f/2} \mbox{ with } N_f \mbox{ fermions } &\leftrightarrow& \quad
USp(2k)_N \mbox{ with } N_f \mbox{ real scalars } .
\end{eqnarray}

None of these dualities have been proven, but they are backed up by a substantial amount of evidence.  First, the field theories are tractable in the 't Hooft limit, that is large $N$ and large $k$ with fixed $N/k$, and in this limit agreement can be verified by the explicit computation of correlation functions and the thermal free energy~\cite{Aharony:2011jz,Giombi:2011kc,Aharony:2012nh,Aharony:2012ns}. This includes a detailed mapping between operators in the two theories. In particular, baryon operators on one side of the duality map into monopole operators on the other side. At finite $N$ one can easily check that the global symmetry charges of these operators match, but in the 't Hooft large $N$ limit one can even match their dimensions \cite{Aharony:2015mjs}. Second, one can show that at any $N$ and $k$ the global symmetries and their `t Hooft anomalies~\cite{Benini:2017dus} match. Third, when adding relevant deformations that drive the theory into massive phases, the resulting topological field theories (TFTs) agree on both sides of the duality \cite{Aharony:2015mjs,Hsin:2016blu}. Last but not least, one can start with a supersymmetric parent duality and deform it in order to trigger a flow to a non-supersymmetric daughter duality. In this way \cite{Jain:2013gza,Gur-Ari:2015pca}, deformed 3d Seiberg duality to flow to 3d bosonization at large $N$. Similarly \cite{Kachru:2016rui,Kachru:2016aon} deformed 3d mirror symmetry to reduce to the non-supersymmetric bosonization duality, at least in the special case\footnote{For this special case, there exists also an exact derivation on the lattice, exhibiting a duality between two lattice gauge theories that are constructed to flow to the bosonic and fermionic side of the duality respectively in the continuum limit \cite{Chen:2017lkr}.} of $N=k=N_f=1$. (This latter deformation comes with some caveats, as explained in~\cite{Kachru:2016rui,Kachru:2016aon}, as the renormalization group flow is not under complete control.) We will discuss these supersymmetric dualities in more detail below.

In this work we present a simple string theory realizations of these 3d bosonization dualities. Embedding field theory dualities into full fledged string theory using brane constructions has a long history starting with the work of \cite{Hanany:1996ie}. These embeddings often help to clarify patterns of the dualities and generate new examples. For supersymmetric gauge theories in three dimensions, two famous dualities can be realized via brane embeddings:

{\it Mirror symmetry} \cite{Intriligator:1996ex}  exchanges the Coulomb branch and Higgs branch of ${\cal N}=4$ supersymmetric Yang-Mills theories. In the brane realizations of \cite{Hanany:1996ie} mirror symmetry is implemented via S-duality in type IIB string theory. Supersymmetry breaking deformations can be systematically added to ${\cal N}=4$ mirror pairs \cite{Tong:2000ky} by gauging the R-symmetry, leading to mirrors with ${\cal N}=2$ supersymmetry. Recently it has even been demonstrated that, at least in the Abelian case, one can flow all the way to the non-supersymmetric seed pair by starting with the supersymmetric mirrors \cite{Kachru:2016rui,Kachru:2016aon}. Mirror pairs with ${\cal N}=2$ supersymmetry have been engineered via S-duality for brane embeddings with rotated branes \cite{deBoer:1996ck,deBoer:1997kr,deBoer:1997ka}. But it is difficult to imagine how one could get the non-supersymmetric bosonization duality out of string theory this way. The embeddings in the style of \cite{Hanany:1996ie} realize the gauge theory on the worldvolume of $N$ ``color" D-branes, with matter being introduced by $N_f$ ``flavor" D-branes. The field theory matter is localized at the brane intersection and arises from strings stretching from color to flavor branes. The type of matter localized on such a brane intersection depends on the number of ND directions, that is the directions that are part of the worldvolume of one brane but not the other. Supersymmetric intersections correspond to 4 ND or 8 ND directions. In the former case one introduces a hypermultiplet of matter with equal number of bosons and fermions. The latter introduces purely fermionic matter, but can only be realized on 0+1 or 1+1 dimensional intersections (both of which allow purely fermionic supermultiplets). To get a 2+1 dimensional gauge theory with fermionic matter one can introduce branes with 6 ND directions as studied in \cite{Davis:2008nv} following the famous Sakai-Sugimoto construction in 3+1 dimensions \cite{Sakai:2004cn}. Scalar matter on the other hand is localized at an intersection with 2 ND directions, albeit with a negative mass squared. Even ignoring the instability of the 2 ND intersections, S-duality changes the type of brane (D5's turn into NS5's, D1's into fundamental strings), but it is hard to see how one could change the number of ND directions via S-duality. So implementing 3d bosonization via S-duality is met with serious obstacles.

For ${\cal N}=2$ an a priori different equivalence is {\it Seiberg duality}. While originally discovered in 3+1 dimensions \cite{Seiberg:1994pq}, Seiberg duality can be compactified to 2+1 dimensions \cite{Karch:1997ux,Aharony:1997gp,Aharony:2013dha}. 3d Seiberg duality becomes particularly rich and interesting when CS terms are included \cite{Giveon:2008zn}. It has been argued in \cite{Jain:2013gza,Gur-Ari:2015pca} that, at least at large $N$, one recovers 3d bosonization by deforming the Seiberg dual pairs of \cite{Giveon:2008zn}. In terms of brane setups, Seiberg duality is also easily implemented, even though its derivation is not as straightforward as simply using S-duality. In the construction of \cite{Hanany:1996ie} of a 2+1 dimensional gauge theory, the color D3 branes are suspended between NS5 branes so that the 3+1 dimensional worldvolume gauge theory lives on an interval and reduces at low energies to the desired 2+1 dimensional gauge theory. The full 3+1 dimensional description in terms of a field theory with boundaries and defects contains a lot of extra data that should be irrelevant for the 2+1 dimensional low energy description: the relative positions of NS5 branes and flavor D5 branes along the compact direction. The authors of \cite{Elitzur:1997fh} identified a particular set of brane moves that takes the original gauge theory into its Seiberg dual by rearranging branes in this extra direction, presumably leaving the low energy theory unchanged. While the original work of \cite{Elitzur:1997fh} dealt with Seiberg duality in 3+1 dimensions, the same EGK moves still apply in 2+1 dimensions as well\footnote{Note that the realizations in terms of branes makes it appear as if Seiberg duality and mirror symmetry are two genuinely unrelated statements. One can however show that at least in some cases there exist a duality of dualities \cite{Jensen:2009xh} where, when lifted to M-theory, a brane setup realizing a mirror pair can be seen to be equivalent to a brane setup realizing a Seiberg dual pair.} \cite{Giveon:2008zn}. Using this strategy one can attempt to derive non-supersymmetric dual pairs as was for example done in \cite{Armoni:2014cia}, but the results look nothing like 3d bosonization. Once again, it is hard to see how the required change of a 6 ND flavor brane to a 2 ND flavor brane could be accomplished by the EGK moves.

Given these difficulties, in this work we will give an embedding of 3d bosonization into string theory via branes using a completely different approach based on flavored holography \cite{Karch:2002sh}. While we will also invoke holography, this realization is quite distinct from the higher spin holographic duals of \cite{Aharony:2011jz,Giombi:2011kc,Aharony:2012nh}. To have a higher spin dual, one needs to take a 't Hooft large $N$ limit, that is large $N$ and $k$ with fixed $N/k$. In particular, both sides of the duality have a large number of colors. In our case, we will appeal to holography at large $N$ with fixed $k$. In this case, the bosonic gauge theory has a small rank, is weakly coupled, and surely has no classical supergravity dual. In our string theory embedding, the fermionic side of the story will be a standard large $N$ gauge theory with $O(N^2)$ degrees of freedom. We engineer this theory in such a way that most excitations are gapped. At low energies we can argue that it reduces to a $SU(N)$ CS matter theory. The full gauge theory has a nice holographic dual. The physical degrees of freedom in this holographic dual are realized very differently: they are fluctuations of supergravity fields as well as those of probe D-branes embedded into the bulk geometry. But of course also in the bulk most of these degrees of freedom will be gapped. After all, for holography to be true the bulk and field theory spectrum has to agree. Most importantly, all fluctuations of the metric give rise to gapped fluctuations, so that at low energy gravity does not contribute any fluctuating degrees of freedom in the bulk. We will see that the only low energy modes in the bulk geometry correspond to the bosonized $U(k)$ theory at level $N$, which arises as a gauge theory living on probe branes in the bulk spacetime. In this way AdS/CFT reduces to the 3d bosonization duality in the low energy limit, with the fermionic CS theory being the low energy limit of the boundary theory whereas the bosonic theory is the low energy limit of the bulk theory. While the identification of the low energy theory in the bulk is fairly straightforward, we have to make some basic assumptions of the strong coupling dynamics on the boundary in order to argue for the desired low energy theory on the fermionic side.

The rest of this note is organized as follows. In the next Section we will review the connection between 3d bosonization and level/rank duality. Following~\cite{Fujita:2009kw} we embed level/rank duality into AdS/CFT. In Section~\ref{S:flavors} we introduce our basic construction, embedding the duality~\eqref{uni} into string theory via a brane embedding. We describe applications of our construction in Section~\ref{S:discussion}. In particular, we recover the $SO$/$USp$ versions of the duality by including orientifolds, and describe how baryons and monopoles are realized on both sides of the string duality.

\section{Level/rank duality}
\label{S:levelrank}

3d bosonization is closely related to level/rank duality \cite{Nakanishi:1990hj,Naculich:1990hg}. Level/rank duality was first introduced as an equivalence between 1+1 dimensional conformal field theories (CFTs) based on WZW models. Using the mapping \cite{Moore:1988qv} between such 1+1 dimensional CFTs and 2+1 dimensional CS gauge theories, level/rank duality can also be read as an equivalence between the latter. The level/rank dualities of interest here are the $N_f=0$ versions of the dualities \eqref{uni} - \eqref{sp}. Since a Chern-Simons gauge field contributes no dynamical degrees of freedom, the field theories appearing in \eqref{uni} - \eqref{sp} at $N_f=0$ are in fact TFTs and not three-dimensional CFTs. These level-rank dualities between the $N_f=0$ theories are an important check of the 3d bosonization dualities, since the latter can be reduced to the former either by setting $N_f=0$ or by following RG flows in phases where all matter fields are massive. As emphasized in \cite{Hsin:2016blu,Aharony:2016jvv}, one can think of the 3d bosonization dualities as ``adding flavors" to level-rank duality. Of course this is not a derivation of bosonization since we are augmenting the theory on both sides, but it motivates the conjectured dualities. The non-trivial part is to argue that fermionic matter on one side corresponds to bosonic matter on the other side.

We implement this idea of adding flavors to level/rank duality using holography. A holographic realization of level/rank duality was given in \cite{Fujita:2009kw} as we now review. The starting point is ${\cal N}=4$ super Yang-Mills theory (SYM) in 3+1 dimensions with gauge group $SU(N)$. In the limit of large $N$ and large 't Hooft coupling $\lambda=g_{YM}^2 N$ this theory has a dual description in terms of type IIB supergravity on AdS$_5\times \mathbb{S}^5$ \cite{Maldacena:1997re}. In order to reduce the theory to a 2+1 dimensional gauge theory at low energies, we compactify the $x_3$ direction on a circle of periodicity $L$ with anti-periodic boundary conditions for the adjoint representation fermions in the ${\cal N}=4$ vector multiplet. These boundary conditions break supersymmetry completely and give all the fermions masses of order $1/L$. Their scalar superpartners, whose mass is no longer protected by supersymmetry, will pick up masses of order $1/L$ from loop corrections. At energy scales below $1/L$ the only degrees of freedom that remain in the theory are the massless gauge bosons. Their $F^2$ kinetic term has a coefficient of order $g_3^{-2} \sim L g_{YM}^{-2}$ and becomes irrelevant at low energies, leaving the gauge bosons without a kinetic term in the infrared (or, formally, they are infinitely strongly coupled). This theory is believed to be completely gapped with no interesting dynamics in the infrared.

In 2+1 dimensions we have another option for a gauge field kinetic term: CS terms. They are marginal and dominate over the $F^2$ term at low energies. In the presence of CS terms the gauge bosons pick up a mass of order $k g_3^2$, leaving behind a topological theory at low energies governed by the CS action. In order to introduce a CS term in the 2+1 dimensional theory we obtained by compactifying ${\cal N}=4$ SYM, we introduce a spatially varying theta angle $\theta(x_3) = 2 \pi k x_3/L$. This linear profile is consistent with the periodicity of the $x_3$ direction for integer $k$ since the theta angle is only well defined up to shifts of $2 \pi$. Upon integrating by parts, the 4d theta term
\beq
S_{\theta} = \frac{1}{8 \pi^2} \int \theta\, \text{tr}(F \wedge F)\,,
\eeq
reduces to a 3d CS term at level $-k$. As an upshot, this compactification of ${\cal N}=4$ SYM on a circle with anti-periodic boundary conditions and a linearly varying $\theta$ angle gives a theory that in the UV is simply ${\cal N}=4$ SYM, but at energies below the compactification scale $1/L$ reduces to a topological $SU(N)_{-k}$ CS theory.

\begin{figure}[h]
\begin{center}
\includegraphics[width=0.65\linewidth]{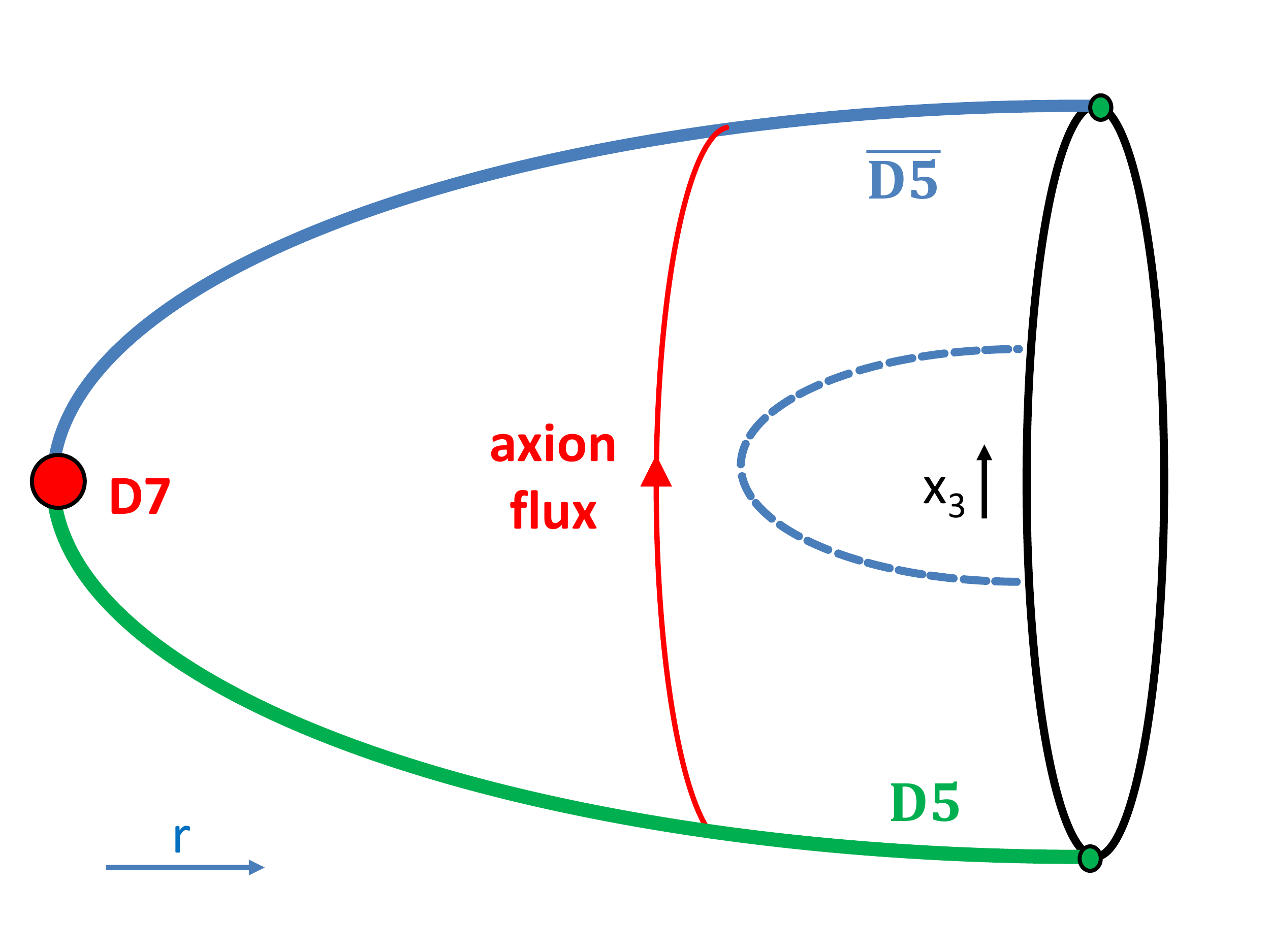}
\end{center}
\caption{
Holographic bulk dual of ${\cal N}=4$ SYM compactified on a circle with anti-periodic boundary conditions for fermions and a linear theta angle. Not displayed are the internal $\mathbb{S}^5$ and the 3d Minkowski part of the spacetime. The $N_f$ flavor D5 and anti-D5 branes extending from the boundary to the Hall D7 branes at the tip of the cigar are introduced in Section~\ref{S:flavors}. The dashed line corresponds to an alternate D5/anti-D5 embedding that is also discussed in Section~\ref{S:flavors}.
}
\label{fig:cigar}
\end{figure}

Let us see how the compactification on a circle with antiperiodic boundary conditions for fermions and a linearly growing axion plays out in the dual geometry. The holographic dual for ${\cal N}=4$ SYM on a circle with anti-periodic fermions is well known \cite{Witten:1998zw}. It is given by the doubly-Wick rotated AdS-Schwarzschild geometry, also known as the AdS-soliton. The metric can be written as
\beq
\label{cigar}
g = R^2 \frac{dr^2}{f(r) r^2} + \frac{r^2}{R^2} \left ( - dt^2
+ dx_1^2 + dx_2^2 + f(r) dx_3^2 \right ) + R^2 g_{\mathbb{S}^5}\,,
\eeq
where $R^4/\alpha'^2 = \lambda$ and $f(r)$ is the ``blackening function" $f(r) = 1-r_0^4/r^4$. The 2d cigar geometry spanned by the two coordinates $r$ and $x_3$ is displayed in Figure \ref{fig:cigar}. It has the topology of a disc. For it to smoothly terminate at $r=r_0$ without a conical singularity we require that $r_0$ is fixed in terms of the curvature radius $R$ and the field theory compactification radius $L$ as $r_0=\pi R^2/L$.

The operator $\text{tr}(F \wedge F)$ is dual to the massless bulk axion field, so in order to turn on a linearly varying $\theta$ in the field theory the bulk axion $\chi$ has to approach $\chi \sim \theta(x_3) = 2 \pi k x_3/L$ at large $r$. Note that
\beq
\label{axion}
\chi = 2 \pi k x_3/L\,,
\eeq
corresponds to a constant axion field-strength and is an exact solution to the axion equation of motions in the fixed geometry of \eqref{cigar}. Of course the full type IIB supergravity equations of motion also include Einstein's equations, which require us to change the geometry due to the extra source of energy density from the axion field strength. But this backreaction of the axion on the background geometry can be neglected\footnote{In the standard normalization of the axion field, in which the asymptotic value of the axion is in fact just the Yang-Mills theta angle, we have an axion kinetic term with no powers of the string coupling, whereas Newton's constant scales as $g_s^2 \sim N^{-2}$ so that the backreaction of an axion field strength of order $k$ is suppressed by $(k/N)^2$.} in the ``probe limit" $k \ll N$. A second problem with the solution \eqref{axion} is that the constant axion field strength requires a source at the origin in the $r-x_3$ plane, that is at $r=r_0$. This source is provided by the introduction of $k$ probe D7 branes located at $r=r_0$, as also displayed in Figure \ref{fig:cigar}. These ``Hall" probe branes span the three non-compact field theory directions $t$, $x_1$ and $x_2$ as well as the entire internal sphere. As for the axion field strength, the backreaction of the probe branes can be neglected for $k \ll N$. Their $\mathbb{R}^{2,1}\times \mathbb{S}^5$ embedding is quite distinct from the flavor probe branes that arise in flavored holography \cite{Karch:2002sh}. Flavor branes change the matter content of the field theory; they reach to asymptotic infinity. In contrast, the Hall probe branes that support the axion in the solution of \cite{Fujita:2009kw} are localized in the radial direction. They are simply a source that can support the axion field strength which encodes the spatially varying field theory theta angle.

The field theory was engineered to give $SU(N)_{-k}$ CS theory in the infrared. To understand how this is encoded in the bulk, we need to identify the low energy limit in the bulk. First note that all IIB supergravity modes in the cigar geometry are gapped \cite{Witten:1998zw}. The corresponding mass gap
\beq
m_{KK} \sim \frac{r_0}{R^2} \sim \frac{1}{L}\,,
\eeq
reproduces field theory expectations. So at energies below $1/L$, the theory is dominated by the Hall probe branes. Due to the cigar geometry, scalar fluctuations of the Hall probe branes receive a mass as they need to fluctuate ``up" into the $r$-direction and the same is expected for the fermions. Since the internal $\mathbb{S}^5$ space is compact, the low energy description is a 2+1 dimensional $U(k)$ gauge theory.\footnote{From the point of view of the $\mathbb{R}^{2,1}\times \mathbb{S}^5$ worldvolume of the D7 branes, the mass of these extra modes is simply of order $1/R$, the compactification radius of the $\mathbb{S}^5$. But note that the Minkowski metric on $\mathbb{R}^{2,1}$ has an overall $r_0^2/R^2$ red-shift factor. Rescaling time, and hence energy, so that if the Minkowski metric is properly normalized the energy $1/R$ again corresponds to an energy $m_{KK}$ for the properly normalized energy.} The D7 worldvolume theory includes WZ couplings of the form $A \wedge F \wedge H$ to the $N$ units of background 5-form flux $H$. This induces a CS term of level $N$ for the worldvolume $U(k)$ theory. Correspondingly, the probe brane theory is also gapped, as it has to be to agree with the field theory. Its low energy limit is simply a topological $U(k)_N$ CS gauge theory. To conclude, the construction of \cite{Fujita:2009kw} shows that the holographic dual of $SU(N)_{-k}$ CS theory, realized as a low-energy limit of ${\cal N}=4$ on a circle, is in fact its level/rank dual $U(k)_N$ CS theory. In the $k \ll N$ probe limit, the infrared limit of the holographic duality is level/rank duality.

\section{Adding Flavors}
\label{S:flavors}

Adding flavors to holography can be accomplished straightforwardly using flavor probe branes \cite{Karch:2002sh} (see \cite{Erdmenger:2007cm} for a review). Having realized level/rank duality in holography, we can use this strategy to explicitly add flavors to level/rank duality. What we aim to show here is that this yields 3d bosonization as anticipated. We add matter by introducing $n_f$ flavor D5 branes as in \cite{Karch:2000gx}. The D5s intersect the color D3 branes along a 2+1 dimensional defect, as can be seen in Table \ref{branetablea}. We will also add a second stack of $n_f$ anti-D5 branes separated from the D5 branes along the $x_3$ direction.\footnote{This construction is a lower-dimensional version of one of the earliest holographic realizations of QCD \cite{Kruczenski:2003uq}, where D6 branes (along 0123567) gave rise to 3+1 dimensional defect hypermultiplets in a 4+1 dimensional gauge theory (living on D4 branes along 01234) compactified on a supersymmetry breaking circle.}

Let us briefly review the physics of the D5 probes before we compactify the $x_3$ direction and turn on the linear theta angle. The D5 branes are 4 ND flavor branes and so each one adds a whole hypermultiplet's worth of fundamental representation matter to the gauge theory, preserving half of the 16 supersymmetries of ${\cal N}=4$ SYM.\footnote{The anti-D5s alone would give rise to the same matter content, but preserving the opposite half of the supersymmetry.} The resulting gauge theory is ${\cal N}=4$ $SU(N)$ SYM coupled to $n_f$ defect hypers so as to preserve 3d $\mathcal{N}=4$ supersymmetry \cite{Karch:2000gx,DeWolfe:2001pq}. In fact this theory is a prototypical example of a defect CFT. The field theory has a manifest $SU(2)_L \times SU(2)_R$ R-symmetry corresponding to rotations in the 456 and 789 planes respectively. In addition, the flavors can be rotated by a $U(n_f)$ flavor symmetry. In the holographic dual bulk, the probe D5 branes span an AdS$_4 \times\mathbb{S}^2$ worldvolume in the background AdS$_5\times \mathbb{S}^5$ geometry. Writing the metric on the internal $\mathbb{S}^5$ in a form that makes the $SU(2)_L \times SU(2)_R$ symmetry manifest,
\beq
g_{\mathbb{S}^5} = d \chi^2 + \cos^2 \chi \,g_{\mathbb{S}^2} + \sin^2 \chi \,g_{\overline{\mathbb{S}^2}}\,,
\eeq
the D5 branes wrap the first $\mathbb{S}^2$ with metric $g_{\mathbb{S}^2}$. The D5 worldvolume gauge field is dual to the $U(n_f)$ flavor symmetry current.

\begin{table}[h]
\begin{center}
\subfloat[Brane realization of the gauge theory. ]{
\begin{tabular}{c|cccccccccc}
\label{branetablea}
&0&1&2&3&4&5&6&7&8&9 \\
\hline
D3& \bf{x}& \bf{x}&\bf{x}&\bf{x}&o&o&o&o&o&o \\
D5& \bf{x}&\bf{x}&\bf{x}&o&\bf{x}&\bf{x}&\bf{x}&o&o&o \\
\end{tabular}}
\quad \hspace{1cm} \quad
\subfloat[Embedding of the probe branes in the AdS-soliton $\times$ $\mathbb{S}^5$ geometry. ]{
\begin{tabular}{c|cccccc}
\label{branetableb}
&0&1&2&3&$r$&$\mathbb{S}^5$ \\
\hline
D7 &  \bf{x}&\bf{x}&\bf{x}& o& o& $\mathbb{S}^5$ \\
D5 & \bf{x}&\bf{x}&\bf{x}&o& \bf{x}& $\mathbb{S}^2$
\end{tabular}}
\caption{Brane setups realizing the field theory as well as its holographic dual. In a) $N$ color D3 branes intersect $n_f$ flavor D5 and anti-D5 branes. in b) the same $n_f$ flavor D5 branes intersect $k$ Hall D7 branes.}
\end{center}
\end{table}

The supersymmetry-preserving mass terms are triplets under $SU(2)_R$; turning them on breaks the R-symmetry to $SU(2)_L \times U(1)$. In the bulk geometry the triplet mass deformation causes the D5 brane to slip off the internal $\mathbb{S}^2$, that is $\chi(r)$ will develop a non-trivial profile. This ``massive" D5 brane smoothly terminates at a finite radial coordinate $r_*$ by contracting at the pole of the internal sphere. The undeformed theory corresponds to the $SU(2)_R$-invariant embedding $\chi(r)=0$. For generic embeddings, the fall-off of $\chi(r)$ near the boundary encodes the dual field theory data. The coefficient of the leading $r^{-1}$ term gives the triplet mass, the coefficient of the subleading $r^{-2}$ gives the corresponding fermion bilinear condensate. The requirement that the brane terminates smoothly at $r_*$ fixes the condensate in terms of the mass. In the full AdS$_5$ $\times$ $\mathbb{S}^5$ geometry these massive embeddings can be found analytically \cite{Karch:2002sh}; they are given by $\sin \chi = m/r$.

As we will discuss in more detail below, the field theory also allows a supersymmetry breaking $SU(2)_R$ singlet mass. This operator is not dual to a light supergravity or brane fluctuation, but rather to a true stringy mode. It will play an important role below.

The task at hand is to understand what happens to the extra flavor degrees of freedom once we compactify the $x_3$ direction and turn on the linear theta angle, and correspondingly replace AdS$_5$ with the cigar geometry and turn on the constant axion field strength. The supersymmetry breaking circle compactification will give mass to many of the flavor degrees of freedom. We need to identify the light degrees of freedom left behind, both in the holographic bulk as well as in the boundary field theory.

\subsection{Bulk Theory}

The correct low energy physics of the bulk theory is easier to understand, and so we will start with this side of the duality. Once we replace AdS$_5$ with the cigar geometry of \eqref{cigar}, it is no longer an option to have a single stack of $n_f$ D5 branes with zero triplet mass and condensate.
The geometry itself ends at $r=r_0$. By charge conservation, the probe D5s must terminate at an $r_* \geq r_0$ even for vanishing triplet mass,
leading to a spontaneous breaking of $SU(2)_L \times SU(2)_R \rightarrow SU(2)_L \times U(1)$. The resulting Goldstone bosons correspond to fluctuations of the D5 embedding in the internal $\mathbb{S}^5$ as has been explicitly demonstrated in the closely related case of D6 defects in the D4/D6 intersection \cite{Kruczenski:2003uq}. These $SU(2)_R$-breaking embeddings do not serve our purpose of adding matter to level/rank duality. The only new light degrees of freedom are the Goldstone bosons, which do not carry charge under the CS gauge fields living on the Hall D7 branes. They completely decouple from the TFT living on the Hall D7s. We need a flavor brane that can reach all the way to the tip of the cigar, with new light degrees of freedom arising from the intersection of flavor D5s and Hall D7s.

Here is where the introduction of the stack of anti-D5 branes helps. Instead of terminating at a finite $r_* > r_0$ by slipping off in the internal space, we now can have embeddings with no slipping mode turned on, thereby preserving the full $SU(2)_L \times SU(2)_R$ R-symmetry, where the D5 and anti-D5 smoothly connect in the bulk into a single "U-shaped" D5 brane as displayed in Figure \ref{fig:cigar}. This type of embedding is very familiar from the Sakai-Sugimoto model \cite{Sakai:2004cn}. Since this embedding preserves $SU(2)_R$, all the $SU(2)$ triplet condensates vanish and the only non-vanishing condensate is the non-local singlet $\langle \xi^{\dagger,i} e^{i \int A} \psi_i \rangle$. Here $\xi$ and $\psi$ denote fermions on the D5 and anti-D5 stacks respectively, $i$ is an $SU(2)_R$ doublet label, and a Wilson line inserted between the two fermions runs from one defect to the other and ensures that the non-local bilinear is gauge-invariant. While leaving the R-symmetry unbroken, this condensate breaks the independent $U(1)_B$ baryon number symmetries on the D5 and anti-D5 stacks to their diagonal subgroup. For general $n_f$, the breaking is $U(n_f)\times U(n_f)\rightarrow U(n_f)$.  This breaking is geometrically encoded by the D5s and anti-D5s connecting into a stack of $n_f$ smooth coincident branes with a single $U(n_f)$ worldvolume gauge field but two asymptotic regions. Unlike the $SU(2)_R$ breaking condensates, which were encoded in the asymptotic fall-off of a bulk field (the slipping mode), the non-local operator that condenses here is not encoded in a local bulk field. Since it involves a Wilson line, it is dual to a string worldsheet. The relevant worldsheet has the topology of a disc with the string endpoints lying on the U-shaped D5, as discussed for the related case of the Sakai-Sugimoto model in \cite{Aharony:2008an}. The area of the worldsheet computes the condensate.

The separation of the two stacks along the $x_3$ circle determines how far down the U-shaped embedding reaches. The generic case is depicted by the dashed "alternate" embedding in Figure \ref{fig:cigar}: the turning point appears at a value $r_t > r_0$ strictly above the tip of the cigar. Since the $U(k)_N$ CS gauge theory lives on the Hall D7 branes localized at the tip of the cigar, these generic embeddings with $r_t >r_0$ once again do not add any light charged matter to the CS gauge theory. Just like the gravity modes, most of the worldvolume modes on the U-shaped D5 brane are gapped by the cigar geometry. The only exception are the Goldstone bosons from the $U(n_f)\times U(N_f) \rightarrow U(n_f)$ breaking associated to the smooth connection of the two brane stacks. They are expected to be encoded in the worldvolume $U(n_f)$ gauge field as in the Sakai-Sugimoto model. But once again they only give us neutral light fields decoupled from the CS theory and do not serve our purpose of adding fundamental flavors to level/rank duality. Said another way, the low-energy theory is a $U(k)_N$ TFT and a decoupled set of Goldstone bosons, not a CS-matter theory.

In the special case of maximal separation $L/2$ between the stacks, the D5s reach all the way down to the bottom of the cigar,\footnote{In formulae, the D5 embedding is characterized by a function $x_3(r)$, in terms of which the area functional is given by (setting $R=1$ for this calculation) $${\cal L} \sim r^2 f^{-1/2} \sqrt{1+ r^4 f^2 x'^2}\,. $$ Since $x_3$ does not appear explicitly in ${\cal L}$ we can integrate $$ x_3' = \frac{\sqrt{f(r_t)} r_t^4}{f(r) r^2  \sqrt{f(r) r^8 - f(r_t) r_t^8}}\,. $$ This can be integrated for generic $r_t$ to get the brane separation at infinity. Choosing $r_t=r_0$, we get the simple solution $x_3'=0$ which corresponds to the maximally separated case.} as displayed by the solid U in Figure \ref{fig:cigar}. In this case new massless matter is localized at the D5/D7 intersection!

What is the extra light matter? From the embedding as displayed in Table \ref{branetableb} we can see that the D5/D7 intersection is 4 ND, so the localized matter is a $U(k) \times U(n_f)$ bifundamental hypermultiplet. This hypermultiplet lives on the $\mathbb{R}^{2,1}\times \mathbb{S}^2$ intersection. Since we are only interested in matter that has mass less than $1/R$, where $R$ is the radius of the $\mathbb{S}^2$, we only want to keep the zero modes of fields on the internal space. For scalars this is the singlet s-wave, so from the $2 n_f$ complex scalars in the hypermultiplet this gives us exactly $N_f=2n_f$ charged scalars coupled to the $U(k)_N$ CS gauge field. However, there are no fermion zero modes on $\mathbb{S}^2$. On the sphere we only have conformal Killing spinors which give rise to 2+1 dimensional fermions of mass of order $1/R$. So at energies below $1/R$ (which, when accounting for the redshift factor of the $\mathbb{R}^{2,1}$ corresponds to field theory energies below $1/L$ as discussed in the previous Section), the field theory we are left with
in the bulk is a $U(k)_N$ CS theory with $N_f=2n_f$ fundamental scalars (in addition to the decoupled Goldstone bosons). This is exactly the field content of the bosonic side of the 3d bosonization pair \eqref{uni}, albeit for even $N_f$.

Our work is not yet done. The bosonic side of the duality~\eqref{uni} is also accompanied by a $U(N_f)=U(2n_f)$-invariant potential, tuned to criticality. In our brane construction, while there are $N_f = 2n_f$ flavors, only a $U(n_f)\times U(n_f)$ flavor symmetry is manifest. This indicates that our construction either realizes the bosonic side of~\eqref{uni} deformed by a potential which breaks $U(2n_f)\to U(n_f)\times U(n_f)$, or the manifest symmetry is enhanced to $U(2n_f)$ in the infrared. Both cases are interesting. In the former, we would end up with a deformation of the original duality~\eqref{uni}, and in the latter, we would find the duality on the nose.

\subsection{Boundary Theory}

To understand the low energy limit on the field theory side, we must understand the strong coupling dynamics of the gauge theory, and so we are on somewhat shaky footing. As with the adjoint scalars of the ${\cal N}=4$ SYM vector multiplet, the hypermultiplet scalars are expected to all become massive via loop corrections since they are unprotected by symmetries. The only candidates for light matter are the fermions and Goldstone bosons. Once again, we are not interested in the latter since they do not correspond to charged matter. In both the bulk and boundary they yield a decoupled sector. So the real question is whether light fermions survive in the field theory.

Naively one might expect fermion masses to be protected by symmetry. In the ultraviolet our field theory has a $SU(2)_L \times SU(2)_R \times  U(1)_{\psi} \times U(1)_{\xi}$ symmetry. Here we focus on the $n_f=1$ case for illustration and the two Abelian factors are the two $U(1)_B$ symmetries. The fermions are neutral under $SU(2)_L$ and so it does not play any role in the discussion. If this whole symmetry were unbroken, fermion mass terms would indeed be strongly constrained. Let us group the fermions $\psi_i$ and $\xi_i$ into a single 4-component fermion $\Psi_I$. We can write the most general mass term invariant under the diagonal $U(1)_B$ symmetry as
\beq
{\cal L}_{m} = \Psi^{\dagger}_I M^{I}{}_J \Psi^J\,,
\eeq
where $M^{I}{}_{J}$ is a Hermitian 4 by 4 matrix. $SU(2)_R$ invariance demands that the triplet masses vanish and so $M$ takes the form
\beq
\label{massmatrix}
M^{I}{}_{J} = \left ( \begin{array}{ll} m_1 & m_2 \cr m_2^* & m_3 \end{array} \right )\,,
\eeq
where each entry is multiplied by a $2\times 2$ identity matrix. An unbroken $U(1)_{\chi} \times U(1)_{\xi}$ symmetry would set $m_2=0$. Additionally, if our theory was time-reversal invariant, that would set $m_1=m_3=0$ since these terms are odd under time-reversal.

But time reversal is not a symmetry once we turn on the linearly varying $\theta$ angle. Furthermore, $U(1)_{\chi} \times U(1)_{\xi}$ is spontaneously broken by the chiral condensate in the U-shaped embedding. So the real question is how many massless fermions remain at energies below the scale of chiral symmetry breaking.

The U shaped embeddings we are interested in preserve $SU(2)_R \times U(1)_B$, so we know that the effective mass matrix in the infrared has to take the form \eqref{massmatrix} with generic real $m_1$ and $m_3$ and complex $m_2$. The eigenvalues of the mass matrix \eqref{massmatrix} come in pairs and are given by
\beq
2 m_{\pm} = m_1 + m_3 \pm \sqrt{(m_1 - m_3)^2 + 4 |m_2|^2}\,.
\eeq
So for generic masses we have no massless fermions, but for specially tuned cases we can have $2 n_f$ or even $4 n_f$ massless fermions. To have $2n_f$ light fermions we must tune one coupling, $|m_2|^2 = m_1 m_3$, whereas in order to have $4n_f$ light fermions all of the effective mass terms to vanish. For the case of $2n _f $ massless fermions we see that the two massive fermions have both positive mass and so, in this phase, the CS level of the $SU(N)$ gauge theory is shifted from $-k$ to $-k + n_f$.

At this point we have to appeal to the holographic dual to determine which of these options is the correct number of light fermions picked by the dynamics of the theory. It is reassuring to note that among the three options we have, $2n_f$ massless fermions together with $2n_f$ positive mass fermions gives us exactly the theory that 3d bosonization demands, $SU(N)_{-k + N_f/2}$ CS gauge theory with $N_f=2n_f$ massless fermions. So this has to be the right option.

Here we give some additional evidence that this is the correct identification by studying massive deformations of the CFT. First of all note, that for generic separations of the D5 and anti-D5 stacks, the U-shaped D5 brane embedding in the bulk does not intersect the Hall D7 branes at all. This is consistent with the fact that we have to tune one parameter in order to get 2$n_f$ light fermions. The extra light matter only arises for maximal separation of the stacks.\footnote{There is also an approximate $\mathbb{Z}_2$ symmetry at maximal separation. It is the combination of $PT$ and an exchange of the D5s and anti-D5s, and it sets $m_3=m_1$. This symmetry is only approximate to $O(1/N)$ and is broken by the linear theta angle, but it is exact for certain values of D5 and anti-D5 locations, like $x=0$ and $x=L/2$ when $k$ is even.} For generic separation the bulk low energy physics remains $U(k)_N$ even in the presence of the flavor D5 branes. This phase appears to correspond giving our $N_f=2n_f$ light fermions a negative mass, cancelling the CS contribution of the $2n_f$ positive mass fermions we already accounted for above and leaving a $SU(N)_{-k}$ CS TFT at low energies.

If our field theory identification is correct, we should also be able to deform our CFT with $N_f=2n_f$ massless fermions with positive masses for the remaining fermions, ending up with a $SU(N)_{-k+N_f}$ CS gauge theory in the infrared. Here we show that it is indeed possible to describe this phase from the probe brane side as well. In order to reach this phase we can turn on the $SU(2)_R$ breaking triplet masses in the UV. While we can not follow explicitly the RG flow in the field theory to see how this affects the effective IR fermion masses, we can analyze the dual probe brane embedding to confirm that it describes an $SU(N)_{-k+N_f}$ topological field theory, or more precisely its $U(N_f-k)_N$ level/rank dual.

Turning on the triplet mass corresponds to turning on a non-trivial boundary condition for the slipping mode $\chi(r)$. There are two qualitatively different embeddings that one could imagine: the D5 branes either slip off the internal sphere at $r_* > r_0$, or they reach $r_0$ at a finite angle and either end on the D7s or meet up with the anti-D5s. In parlance common in the probe brane literature we call the former ``Minkowski embeddings.'' It is easy to confirm from a numerical analysis of the worldvolume action that for small triplet masses the Minkowski embedding has larger action and only becomes the preferred configuration above a finite triplet mass $m_*$. The transition to the Minkowski embeddings is first order. We give the details of the corresponding probe brane analysis in the Appendix. For us this means that as far as the embeddings with small mass are concerned, which describe the possible deformations of our CFT, we can completely ignore the Minkowski embeddings. The only embeddings of interest are the ones where the D5 and anti-D5 probe branes reach the tip of the cigar at $r_0$. If we give both D5 and anti-D5 the same triplet mass, the probes can still reconnect into a U-shape, even though now at the bottom of the U they no longer wrap an equatorial $\mathbb{S}^2$ but a slightly smaller $\mathbb{S}^2$. Nevertheless, this does not change the low energy physics, which remains $U(k)_N$ with $N_f$ scalars. The hypermultiplet is still living on $\mathbb{R}^{2,1}\times \mathbb{S}^2$. The $\mathbb{S}^2$ radius is reduced compared to the massless embedding, but this does not change the physics of the scalar zero-mode. So at the face of it it appears that this finely tuned choice of equal D5 and anti-D5 masses is irrelevant from the point of view of the low energy physics.

For generic small triplet masses, the D5s and anti-D5s have to end on the Hall D7 branes since they hit it at different locations. We can think of them $n_f$ U-shaped D5s that split on the D7s. Splitting all of the flavor D5s across the Hall D7s takes us to a massive phase. More precisely, this exactly corresponds to giving the hypermultiplet living at the D5/D7 intersection a vacuum expectation value. As there are $N_f=2n_f$ charged scalars, the $U(k)_N$ theory can be Higgsed down to $U(k-N_f)_N$, at least assuming the Higgsing is maximal.\footnote{The maximal Higgsing recalls the S-rule of~\cite{Hanany:1996ie}, which, regrettably, does not apply here as supersymmetry is completely broken. Nevertheless we assume that this symmetry breaking pattern is energetically preferred.} The global $U(n_f)$ flavor symmetry in this phase is given by a linear combination of the original $U(n_f) \times U(n_f)$ flavor and the broken $U(2nf)$ gauge generators. This color-flavor locking leads to a flavor CS term in the low-energy theory, that is a contact term for background gauge fields associated with this global flavor symmetry. The resulting flavor CS term is exactly what one would get in the dual description from giving $N_f$ fermions positive masses. The fact that we can find this phase in the holographic dual gives us great confidence that our identification of $2n_f$ as the right number of massless fermions in the field theory description is correct. Certainly the existence of this phase is inconsistent with having no light fermions. In principle we could also obtain this phase by starting with $4n_f$ fermions and giving $3n_f$ a positive mass and $n_f$ negative mass. But given that we only tuned one parameter (the stack separation), whereas $4n_f$ massless fermions requires four tunings (since all four real mass parameters need to vanish), it is much more natural to conjecture that we indeed have $2n_f$ massless fermions.

Lo and behold, we see that holographic duality at low energies in this particular geometry reduces to 3d bosonization. The $SU(N)_{-k+N_f/2}$ gauge theory with $N_f$ fermions arises as the low energy limit of the boundary field theory, whereas the $U(k)_N$ gauge theory with $N_f$ bosons describes the only light degrees of freedom of the dual string and probe brane description.

\section{Applications and Discussion}
\label{S:discussion}

Having successfully embedded 3d bosonization dualities into string theory, we would like to use our construction to gain new insights. The intuitive geometric nature of the string theory construction should make it easy to understand new deformations and generalizations of the duality. In this section we present two applications where we use the string theory construction to re-derive generalizations of 3d bosonization to $SO$/$Sp$ groups, as well as the details of the operator mapping. We also comment on potential future applications.

\subsection{Orientifolds}

In the previous Section we
embedded the unitary 3d bosonization duality \eqref{uni} into string theory as a holographic duality. To similarly obtain the real dualities \eqref{so} and \eqref{sp} we can augment the construction with orientifolds. The most natural idea is to superpose an orientifold O5 plane on top of the U-shaped D5-branes. We have two choices, O5$^-$ and O5$^+$, which give rise to an $SO(n_f)$ or $USp(2n_f)$ gauge group on the D5 brane respectively. These symmetry groups appear as a global flavor symmetry in our duality. The real question is what is the resulting gauge group on the D3 branes (which started out with an $SU(N)_{-k}$ before orientifolding) and the D7 branes (which realized the $U(k)_N$). Fortunately this analysis is simple, since both the D3 and the D7 have a relative number of 4 ND directions with the D5s. In this case it has been well known e.g from the D1-D5-D9 system of \cite{Douglas:1995bn} that we obtain the opposite projection for the D3s and D7s than we do for the D5s. That is, for an $USp(2n_f)$ flavor symmetry we obtain $SO(N)_{-k}$ and $SO(k)_N$ on the color D3s and Hall D7s respectively, whereas for the $SO( n_f)$ flavor symmetry the gauge groups are $USp(2N)_{-k}$ and $USp(2k)_N$. The D5 matter branes again add purely bosonic matter in the bulk and fermionic matter to the boundary field theory. Assuming that the mass matrix has the same pattern as in the unitary case, we again give masses to half of the fermions, which gives the extra shift in the Chern-Simons level to exactly reproduce the dualities \eqref{so} and \eqref{sp} with $N_f = 2n_f$.

\subsection{Baryons and Monopoles}

One of the more interesting aspects of 3d bosonization is that it maps baryon number on one side of the duality to monopole number on the other side. On the field theory side, which gave rise to $SU(N)_{-k+N_f/2}$ CS theory with $N_f$ fermions, we have a $U(1)$ global symmetry that is simply baryon number. It acts on the matter fields by an overall phase rotation. In the bulk low energy $U(k)_N$ CS theory with $N_f$ scalars, baryon number is gauged and so it is not a physical global symmetry. All physical states are neutral under it. Instead we have a $U(1)$ monopole number whose identically conserved current is $j=\frac{*\text{tr}(F)}{2\pi}$.

How can we see that operators charged under baryon number turn into monopoles under the duality? In the field theory, the fundamental fermions $\psi$ carry charge under the $U(1)$, but the only gauge invariant operator that carries $U(1)$ baryon charge is, as the name suggests, the baryon $\sim \psi \psi \hdots \psi$, built from $N$ fundamental fermions with color indices antisymmetrized. It is well known that the bulk dual of a baryon is~\cite{Witten:1998xy} a baryon vertex, that is, a D5 brane wrapping the entire $\mathbb{S}^5$ sitting at the bottom of the cigar and at a point in $\mathbb{R}^2$. By charge conservation the baryon vertex requires $N$ strings attached. Without flavor branes these strings would have to run out to the boundary and give the baryon an infinite mass. But for us these strings can end on the probe D5's, so that the dual operator is indeed a baryon built from the defect fermions coming from the D5 and anti-D5 branes.

The baryon vertex is exactly what we need to realize a monopole in the bosonic dual. The simplest monopole operators in a $U(k)_N$ Chern-Simons theory are a single unit of magnetic flux accompanied by $N$ fundamental fields, so that the whole operator is gauge-invariant. Now, note that the baryon-vertex D5 is completely embedded inside the Hall D7s. The baryon D5 and Hall D7s form a 2 ND system. Correspondingly they attract each other and the D5 dissolves itself inside the D7s as a unit magnetic flux. So the baryon D5 is a monopole operator in the D7-brane gauge theory.

This matching of monopoles and baryons can be further solidified by looking at all the quantum numbers. Let's take a single D5 vertex since this is the simplest monopole. Once it is dissolved into the Hall D7s, we can pick a $U(k)$ gauge so that it sources one unit of magnetic flux on a single D7; more precisely, its GNO charges are equivalent to $\{ 1,0,0..,0 \}$. The monopole needs $N$ fundamental charges attached to it in order to make a gauge-invariant operator; in the bulk these are $N$ scalars provided by the $N$ strings ending on the vertex, running between the vertex and the flavor D5s. But the basic scalar harmonic in a monopole background is spin-1/2, and there are $N_f=2n_f$ scalars, so there are really $(2N_f)^N$ different monopoles that one can get this way. As for the dual baryon, $\psi \psi \hdots \psi$, each fermion can be in one of the two spin states and be one of the $N_f$ flavors, also leading to $(2N_f)^N$ such baryons.

\subsection{Future Applications}

We have demonstrated two useful applications of our construction above. There are many further questions for which we believe the stringy embedding will serve as a useful tool:
\begin{enumerate}
\item CS gauge theories in 2+1 dimensions necessarily have
non-trivial edge states by anomaly matching. So it is a very natural question to ask how 3d bosonization operates in the presence of boundaries and defects. Exactly the same holographic realization of pure CS theory via compactified D3 branes we also employed in here has recently been used \cite{Fujita:2016gmu} in order to study defects in 2+1 dimensional CS theories across which the CS level jumps. It would be interesting to see how these defects transform under 3d bosonization once flavors have been added.
\item The simple bosonization dualities discussed in this work involve a single gauge group factor on both sides. By gauging background fields, 3d bosonization dualities involving product gauge groups with bifundamental matter can be constructed \cite{Jensen:2017dso}. String theory may be able to reproduce these more complicated dualities as well and perhaps help organize patterns among them. Relatedly, given that the low-energy theory on a stack of M2-branes at an orbifold singularity is a quiver Chern-Simons matter theory, perhaps these quiver bosonization dualities can also be embedded in M-theory.
\item Compactifying 3d bosonization to lower dimensions should allow one to derive new dualities or make contact with known ones as has been demonstrated in the supersymmetric case \cite{Aganagic:2001uw}. The string theory embeddings presented in here may provide guiding principles on how to do this when supersymmetry is broken.
\item Our stringy embedding of bosonization relied on realizing a Chern-Simons matter theory as the low energy limit of a full-fledged large $N$ gauge theory with a known holographic dual. Clearly this UV completion is not unique. So there are many other stringy descriptions one could pursue, potentially giving yet another perspective on these dualities. One obvious alternative starting point is the theory living on the worldvolume of $N$ D4 branes. We can compactify two directions on circles, one with periodic and one with anti-periodic boundary conditions. The dual geometry has an internal $S^4$ and an $M_{2,1}$ fibered over a 3d geometry that has the topology of a cigar times a circle. After the first circle compactification we can view the field theory as a 3+1 dimensional gauge theory whose theta angle is given by the Wilson line of the RR 1-form. Turning on a spatially varying $\theta$ angle hence means we are turning on an RR 2-form field strength $F_{12}$, where $12$ denotes the two circle directions. In the probe limit, the backreaction of the constant $F_{12}$ can be neglected, but D6 brane sources wrapping the internal $S^4$ have to be introduced at the tip of the cigar. The low energy physics of this type IIA construction appears identical to the IIB description we discussed here. It would be interesting to see if with similar constructions we can realize more dualities of ``flavored topological field theories,'' maybe in dimensions other than 2+1.

\end{enumerate}

\section*{Acknowledgements}

We would like to thank Francesco Benini and Nati Seiberg for useful discussions. The work of KJ and AK was supported in part by the US Department of Energy respectively under grant numbers DE-SC0013682 and DE-SC0011637.

\begin{appendix}
\section{Numerical analysis of a D5 probe brane on the cigar geometry}
\label{A:numerics}

In this Appendix we solve the equations of motion for a single D5 brane in the cigar geometry \eqref{cigar}. We consider both ``Minkowski embeddings" where the D5 smoothly ends at a finite $r_* > r_0$ by slipping off the internal sphere, as well as ``black hole" embeddings where the D5 brane reaches the tip of the cigar. The latter is really a misnomer in our case, but follows the standard naming conventions in the probe brane literature. The tip of the cigar at $r=r_0$ is not a Euclidean horizon but rather the bottom of the cigar. A D5 brane cannot simply end there unless there is an object for it to end on. As discussed in the main text, our setup has $k$ D7 branes located at the tip of the cigar, and indeed, D5s can end on D7s. So ``black hole'' embeddings are consistent. The ``Minkowski embeddings'' correspond to a phase where the low energy theory is a TFT and a decoupled sigma model for the Goldstone bosons of the $U(n_f)\times U(n_f)\to U(n_f)$ symmetry breaking, while the ``black hole'' embeddings correspond to a gapless phase with light modes coming from the 5-7 strings. We refer to these as the ``topological'' and ``gapless'' phases. (Strictly speaking, both phases have a decoupled sigma model and so are gapless.)

By solving the equations of motion for the slipping mode $\chi(r)$ we want to map out the phase diagram of the theory. That is, determine the range of masses for which the embeddings exist. If more than one embedding is possible for a given mass we need to find out which one corresponds to a lower free energy. To do so, we need to evaluate the regulated, on-shell, Euclidean action of the D5 probe. We find that for large masses only Minkowski embeddings exist. This is to be expected; in AdS$_5\times \mathbb{S}^5$ only Minkowski embeddings exist and the effects of the blackening function defining the cigar become negligible at large mass. Below a critical mass, $m_*$, black hole embeddings become possible while Minkowski embeddings also still exist. Somewhat surprisingly we find that, for masses below $m_*$, the dominant solution is always a black hole embedding. Correspondingly the transition from the topological phase to gapless phase is first order and happens at $m_*$.

To find the embeddings we need to solve the equations of motion for the slipping mode. Since we do not turn on any worldvolume gauge fields, the action for the slipping mode is simply the area of the D5 brane. The corresponding Lagrangian density is
\beq
{\cal L} =-T r^2 \cos^2 \chi \, \sqrt{f^{-1}(r) + r^2 \chi'^2}.
\eeq
$T$ is the effective brane tension; it is equal to the standard D5 tension times the volume of the internal $\mathbb{S}^2$. The mass and condensate are given by the coefficient of $r^{-1}$ and (minus $T$ times) the coefficient of $r^{-2}$ in the asymptotic expansions of $\chi(r)$ at large $r$. For Minkowski embeddings we require that the embedding smoothly truncates at a finite $r_*$ by reaching $\chi(r) = \pi/2$. The absence of a conical singularity requires that $\chi'(r)$ goes to infinity at this point. This picks a unique solution for a given $r_*$ and so fixes the condensate as a function of mass. For the black hole embeddings we require that the D5 brane ends orthogonally on the D7 brane. This also picks a unique solution for a given mass.

\begin{figure}
        \centering
        \subfloat[Condensate as a function of mass. \label{fig:condensate}]{\includegraphics[scale=0.53]{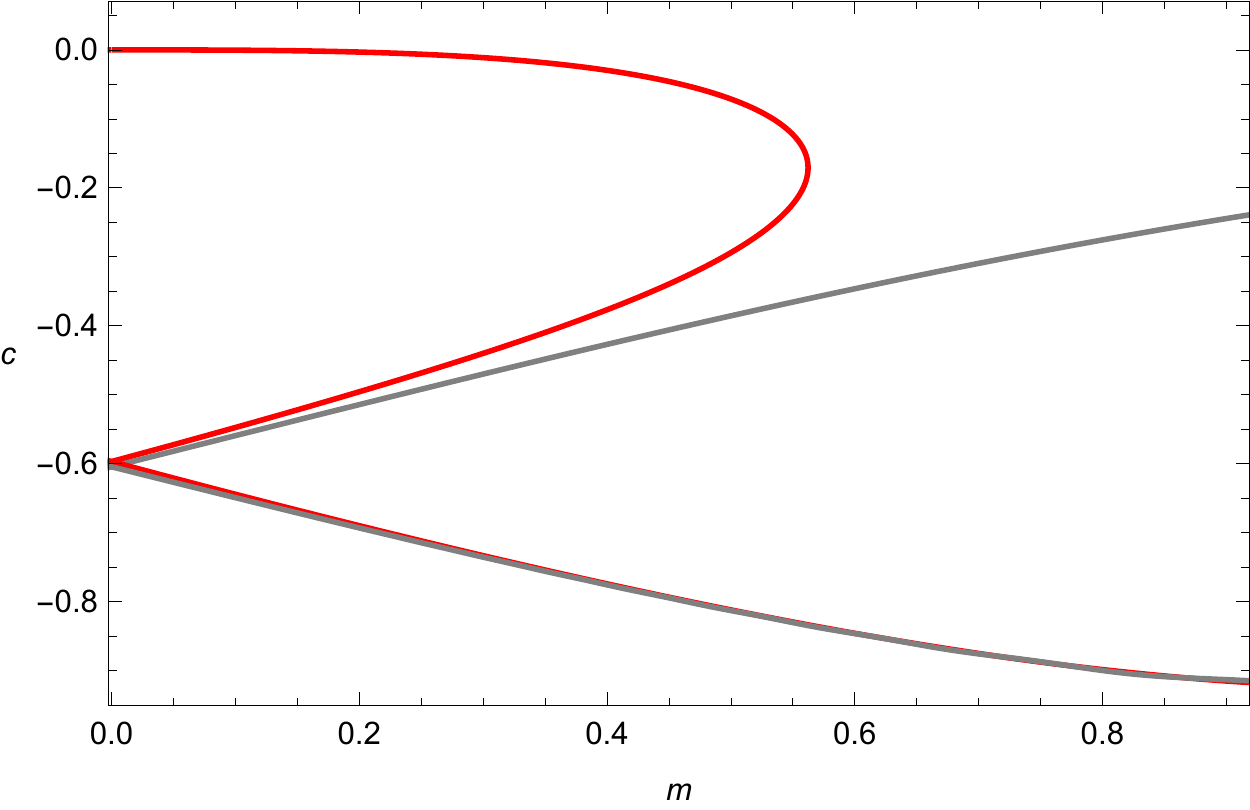}}
        \hskip50pt
        \subfloat[Free energy as a function of mass. \label{fig:freeenergy}]{\includegraphics[scale=0.53]{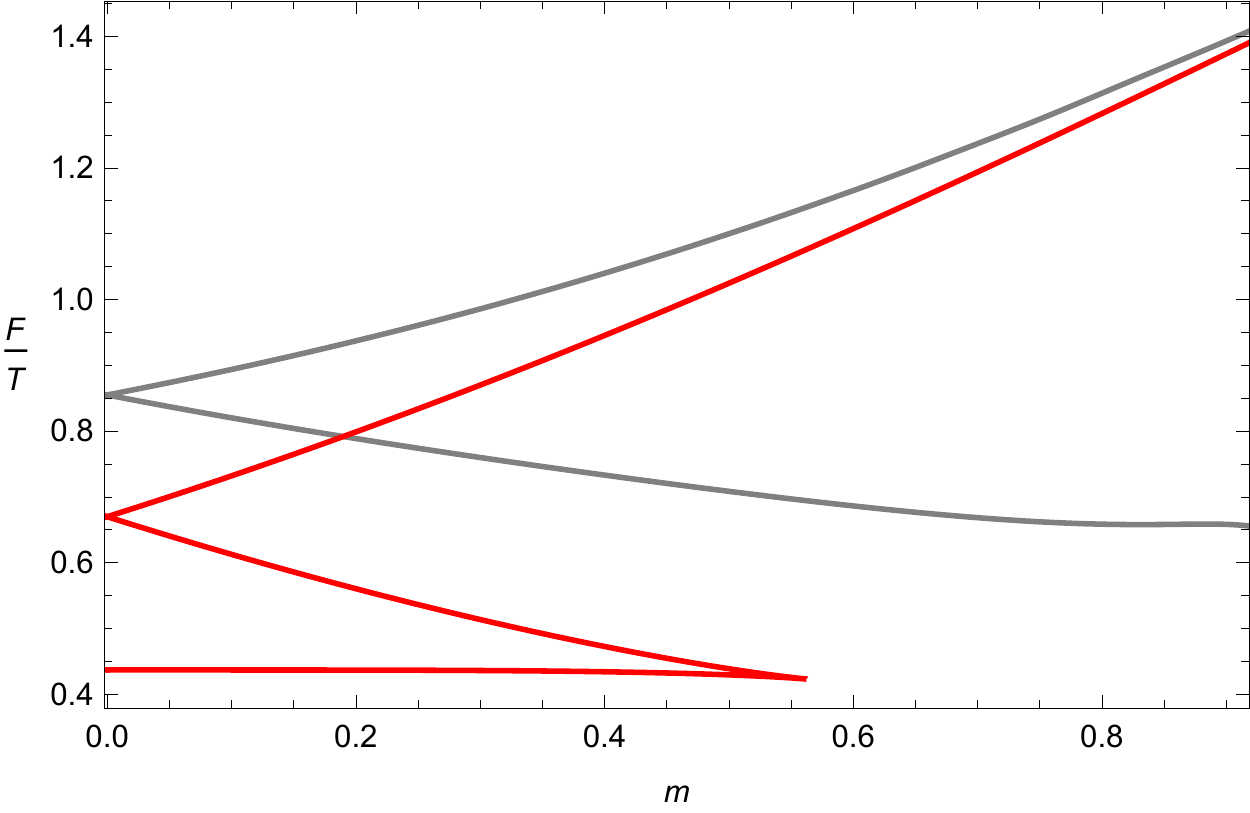}}
		\caption{The $SU(2)_R$ triplet condensate and free energy as a function of triplet mass in the topological (black) and gapless (red) phases respectively. The gapless phase dominates the topological one whenever it exists. We work in units where the parameter $r_0$ appearing in~\eqref{cigar} is unity.
		} \label{fig:probes}
\end{figure}

Our results are displayed in Figure \ref{fig:probes}. Figure \ref{fig:condensate} gives the condensate as a function of mass. We see that above a mass $m_*$ of about\footnote{Strictly speaking, when we are referring to mass here, we simply display the coefficient of the leading $r^{-1}$ fall-off in $\chi(r)$. This gives the flavor mass in units of $1/L$, but to be precise one also needs an extra conversion factor of $\sqrt{\lambda}/(2 \pi)$. This will not be important for us here so we ignore these conversion factors in what follows.} 0.5 only Minkowski embeddings exist, but for masses below this both Minkowski and black hole embeddings are possible.

To determine which configuration is the preferred one we want to determine the free energy corresponding to the two possible solutions. For this we simply need to calculate the Euclidean on-shell action. As usual, the on-shell action is infinite but can be regulated and renormalized by integrating the action up to a ``cut-off slice'' at large radius and adding suitable local counterterms. The counterterms have been worked out for the D5 brane embeddings in \cite{Karch:2005ms}. Figure \ref{fig:freeenergy} shows that as soon as the black hole embedding exists, its corresponding free energy is always below that of the Minkowski embedding. So in this range of masses we can always ignore the Minkowski embeddings. The true ground state of the system is given by the black hole embedding, and the system has a transition at $m=m_*$ to the gapless phase.

\bibliographystyle{JHEP}
\bibliography{stringmirror_revamped}

\end{appendix}

\end{document}